\documentclass[preprint,aps,showpacs,amsfonts,epsf]{revtex4}

\input{epsf.tex}

\newcommand{\E}{{\cal{E}}}

\newcommand{\wb}{\mbox{\boldmath$\omega$}}

\renewcommand{\a}{\alpha}

\newcommand{\dfrac}[2]{\displaystyle\frac{#1}{#2}}
\newcommand{\be}{\begin{equation}}
\newcommand{\ee}{\end{equation}}
\newcommand{\bea}{\begin{eqnarray}}
\newcommand{\eea}{\end{eqnarray}}
\newcommand{\ba}{\begin{array}}
\newcommand{\ea}{\end{array}}

\def\J#1#2#3#4{{#1} {#2} (#3) #4}
\def\PRD{Phys. Rev. D}
\def\PR{Phys. Rev.}
\def\PRL{Phys. Rev. Lett.}

\def\APL{Ann. Phys. (Leipzig)}

\def\PLB{Phys. Lett. B}

\begin{document}
\draft
\title{On electromagnetic energy in Bardeen and ABG spacetimes}

\author{V.S. Manko$\dag$ and E. Ruiz$\ddag$}
\address{$\dag$Departamento de F\'\i sica, Centro de Investigaci\'on y de
Estudios Avanzados del IPN, A.P. 14-740, 07000 Ciudad de M\'exico,
Mexico\\ $\ddag$Instituto Universitario de F\'{i}sica Fundamental
y Matem\'aticas, Universidad de Salamanca, 37008 Salamanca, Spain}

\begin{abstract}
We demonstrate that the total energy of electromagnetic field in
the Bardeen and Ay\'on-Beato-Garc\'ia singularity-free models is
equal to the mass parameter $M$, being therefore independent of
the charge parameter $Q$. Our result is fully congruent with the
original idea of Born and Infeld to use nonlinear electrodynamics
for proving the electromagnetic nature of mass.
\end{abstract}


\maketitle


\section{Introduction}
\label{}

The construction of regular black hole models was pioneered by
Bardeen \cite{Bar} who ingeniously modified the well-known
Reissner-Norstr\"om metric \cite{Rei,Nor} for spherically
symmetric mass and charge to remove the singularity at $r=0$.
Though the global mathematical properties of Bardeen's spacetime
are well understood (see, e.g., \cite{Bor}), the corresponding
electromagnetic source for this spacetime was unknown for many
years, as the Bardeen model did not originally arise as a solution
to some field equations. The first exact regular black hole
solutions were constructed, within the framework of Einstein's
gravity coupled to nonlinear electrodynamics, by Ay\'on-Beato and
Garc\'ia \cite{AGa1,AGa2} who also later reinterpreted Bardeen's
model as an exact solution for a nonlinear magnetic monopole
\cite{AGa3}. Despite a considerable attention these solutions have
received in recent years, it seems that the main physical question
concerning the Bardeen and ABG spacetimes -- How can an
arbitrarily small charge remove the physical Schwarzschild
singularity of a collapsed star with enormous mass? -- still has
not been clarified so far. Being strongly convinced that the
answer to this question must be closely related to the issue of
electromagnetic energy associated with the above spacetimes, in
the present letter we will calculate the total electromagnetic
energy in the Bardeen and ABG models to reveal that for all these
models it has the same value that does not actually depend on the
charge parameter $Q$.

\section{The total energy of electric field in ABG solutions}

We start our consideration with the first ABG solution \cite{AGa1}
defined by the metric
\be d s^2=-fdt^2+f^{-1}dr^2+r^2d\Omega^2, \quad
f=1-\frac{2Mr^2}{(r^2+Q^2)^{3/2}}+ \frac{Q^2r^2}{(r^2+Q^2)^{2}},
\label{AG1} \ee
and the associated electric field
\be E=Qr^4\left(\frac{r^2-5Q^2}{(r^2+Q^2)^4}+\frac{15}{2}
\frac{M}{(r^2+Q^2)^{7/2}}\right), \label{EP1} \ee
the parameters $M$ and $Q$ standing for the mass and electric
charge of the source, respectively.

For a static observer, $u^\a=(-g_{tt})^{-1/2}\xi^\a$, $\xi^\a$
being the timelike Killing vector, the energy density of the
electromagnetic field is defined as
\be T_{\a\beta}u^\a u^\beta=(-g_{tt})^{-1}T_{tt}=-T^t_t,
\label{de} \ee
where $T_{\a\beta}$ is the electromagnetic energy-momentum tensor,
so that the quantity $-T^t_t$ can be calculated either via the
construction of the tensor $T_{\a\beta}$ from the corresponding
tensor $F_{\a\beta}$ of nonlinear electrodynamics or, more
directly, from the Einstein equations,
\be T_{\a\beta}=\frac{1}{8\pi}G_{\a\beta}. \label{Ee} \ee
The total electric or magnetic energy, which is of interest to us
in this letter, then will be equal to the integral over the
surface $t={\rm const}$,
\be {\cal
E}_{e/m}=\int_{R^3}\left(-T^t_t\right)\sqrt{-g}drd\vartheta
d\varphi, \label{en} \ee
where $\sqrt{-g}=r^2\sin\vartheta$ for all the spherically
symmetric spacetimes to be considered.

Let us first obtain the electric energy density for the ABG
solution (\ref{AG1}) straightforwardly from (\ref{Ee}). Then we
get\footnote{For various analytical calculations involving
tensorial quantities, we used the computer program RICCI
\cite{Agu}.}
\be -T^t_t=-\frac{1}{8\pi}G_{tt}g^{tt}=
\frac{Q^2(r^2-3Q^2+6M\sqrt{r^2+Q^2})}{8\pi(r^2+Q^2)^3},
\label{dAG1} \ee
and it can be shown that the density is a positive definite
function if $2M>|Q|$.

While evaluating the total electric energy of the ABG solution
(\ref{AG1}) by means of formula (\ref{en}), we find it instructive
to carry out the integration over $r$ on the interval $[0,r]$,
thus getting ${\cal E}_e(r)$, and then tend $r$ to infinity.
Therefore, taking into account (\ref{dAG1}), we obtain
\bea {\cal E}_e(r)=\int_0^r\int_0^\pi\int_0^{2\pi}
\left(-T^t_t\right)r^2\sin\vartheta drd\vartheta d\varphi
\nonumber\\ =-\frac{Q^2r^3}{2(r^2+Q^2)^2}
+\frac{Mr^3}{(r^2+Q^2)^{3/2}}, \label{ER1} \eea
whence it is fairly well clear how in the limit $r\to\infty$
vanishes the first term on the right-hand side of (\ref{ER1}),
with $Q$ as a factor, while the second term leads to
\be {\cal E}_e(\infty)=M. \label{EAG1} \ee

Of course, one would come to the same result for ${\cal E}_e$ if
one calculates the component $T^t_t$ not by means of the Einstein
tensor (\ref{Ee}) but directly from the energy-momentum tensor of
electric field defined in \cite{AGa1} as
\be 4\pi T^\a_\beta={\mathcal H}_P P_{\beta\mu}P^{\a\mu}
-\delta^\a_\beta(2P{\mathcal H}_P-{\mathcal H}). \label{Tab} \ee
Indeed, taking into account that for the ABG solution (\ref{AG1})
\bea P_{\a\beta}&=&2\delta^t_{[\a}\delta^r_{\beta]}\frac{Q}{r^2},
\quad P^{\a\beta}=-2\delta_t^{[\a}\delta_r^{\beta]}\frac{Q}{r^2},
\quad P=-\frac{Q^2}{2r^4}, \nonumber\\ {\mathcal
H}_P&=&\frac{r^6(2r^2-10Q^2+15M\sqrt{r^2+Q^2})}{2(r^2+Q^2)^4},
\nonumber\\ {\mathcal
H}&=&-\frac{Q^2(r^2-3Q^2+6M\sqrt{r^2+Q^2})}{2(r^2+Q^2)^3},
\label{HP} \eea
it is easy to check that (\ref{Tab}) and (\ref{HP}) yield the same
expression for the energy density as in (\ref{dAG1}), and
consequently the same value of the total electric energy
(\ref{EAG1}).

To be sure that the parameter $M$ in (\ref{EAG1}) is the ADM mass
\cite{ADM} of the ABG solution, let us consider the Komar
\cite{Kom} mass function $M_K(r)$ defined by the following
integral of the 2-form
$\wb=-\textstyle{\frac{1}{2}}\eta_{\a\beta\nu\mu}
\nabla^\nu\xi^\mu dx^\a\wedge dx^\beta$:
\be M_K(r)=\frac{1}{4\pi}\int_{S_r}\wb, \label{Km} \ee
which represents the ``mass'' inside a sphere of radius $r$, so
that the ADM mass will correspond to $M_K(\infty)$. In the case of
the metric (\ref{AG1}), $\wb$ takes the form
\be \wb=\frac{1}{2}\omega_{\vartheta\varphi}d\vartheta\wedge
d\varphi, \label{wAG1} \ee
with
\be \omega_{\vartheta\varphi} =-\frac{2r^3[Q^2(r^2-Q^2)
-M(r^2-2Q^2)\sqrt{r^2+Q^2}]\sin\vartheta} {(r^2+Q^2)^3},
\label{wtf1} \ee
and thus we have
\bea M_K(r)&=&\frac{1}{8\pi}\int_0^{\pi}\int_0^{2\pi}
\omega_{\vartheta\varphi}d\vartheta d\varphi=
\frac{1}{4}\int_0^\pi\omega_{\vartheta\varphi}d\vartheta
\nonumber\\ &=&-\frac{r^3[Q^2(r^2-Q^2)
-M(r^2-2Q^2)\sqrt{r^2+Q^2}]} {(r^2+Q^2)^3}, \label{KAG1} \eea
whence, in the limit $r\to\infty$, we finally arrive at
\be M_K(\infty)=M. \label{ADM1} \ee

Therefore, the total electric energy of the solution (\ref{AG1})
is equal to the ADM mass $M$ independently of the value of the
charge parameter $Q$. Though this result may look surprising at
first glance, it nevertheless is quite logic as it leaves no doubt
that the electric energy in the metric (\ref{AG1}) is comparable
with the ADM mass and hence seems to be able to regularize the
Schwarzschild singular spacetime in principle. At the same time,
it is also clear that the ABG solution (\ref{AG1}) can hardly
describe the field of a point charge, but rather of some
distribution of positive and negative charges for which the
particular value of $Q$, playing in such a case the role of a net
charge, does not really matter. In Fig.~1 we have plotted the
functions ${\cal E}_e(x)/M$ and $M_K(x)/M$ of this solution versus
the dimensionless variable $x=r/|Q|$. Note also that the total
electric energy corresponding to the ``massless'' ($M=0$)
subfamily of the metric (\ref{AG1}) is zero for any $Q$, which is
an indication that this one-parameter spacetime must have regions
of positive and negative energy.

\begin{figure}[htb]
\centerline{\epsfysize=55mm\epsffile{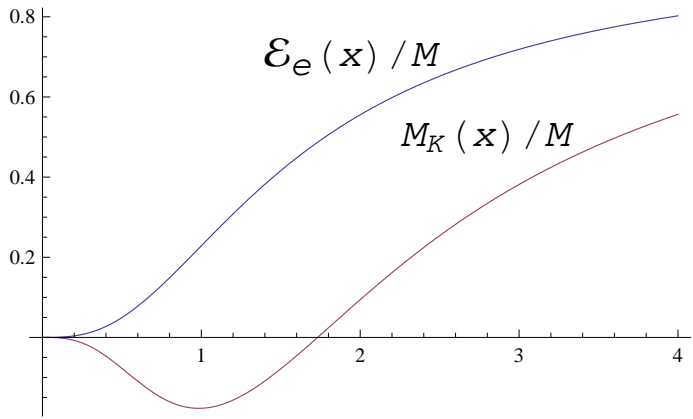}} \caption{Behavior
of the functions ${\cal E}_e(x)/M$ and $M_K(x)/M$, with $x=r/|Q|$
and $|Q|/M=1$, in the case of the first ABG solution.}
\end{figure}

\subsection{The second ABG solution}

It turns out that the above said about the energy of the ABG
metric (\ref{AG1}) is fully applicable to another ABG spacetime
described by the metric \cite{AGa2}
\be d s^2=-fdt^2+f^{-1}dr^2+r^2d\Omega^2, \quad
f=1-\frac{2M}{r}\left(1-\tanh\frac{Q^2}{2Mr}\right), \label{AG2}
\ee
and the electric potential
\be E=\frac{Q}{4Mr^3}\left(1-\tanh^2\frac{Q^2}{2Mr}\right)
\left(4Mr-Q^2\tanh\frac{Q^2}{2Mr}\right). \label{EP2} \ee

Indeed, like in the previous case, the density of electric field
can be evaluated through the Einstein tensor, yielding
\be -T^t_t= \frac{Q^2}{8\pi r^4}\,{\rm sech}^2\frac{Q^2}{2Mr},
\label{dAG2} \ee
which is a positive definite function for any nonzero values of
$M$ and $Q$. Then the electric energy contained inside a sphere of
radius $r$ is given by the expression
\be {\cal E}_e(r)=M-M\tanh\frac{Q^2}{2Mr}, \label{ER2} \ee
and, for large $r$, it behaves as
$M-\frac{Q^2}{2r}+O\left(\frac{1}{r^2}\right)$,
so that for the total energy of electric field $\E_e(\infty)$ we
again obtain, after taking the limit $r\to\infty$ in (\ref{ER2}),
the value $M$.

The Komar mass function $M_K(r)$ of the second ABG solution is
determined by the formulas (\ref{Km}) and (\ref{wAG1}) with
\be \omega_{\vartheta\varphi}
=\sin\vartheta\left[2M\left(1-\tanh\frac{Q^2}{2Mr}
\right)-\frac{Q^2}{r}\,{\rm sech}^2\frac{Q^2}{2Mr}\right],
\label{wtf2} \ee
and therefore, taking into account (\ref{KAG1}), we get
\be M_K(r)=M-\frac{Q^2}{r\left(1+\cosh\dfrac{Q^2}{Mr}\right)}
-M\tanh\frac{Q^2}{2Mr}. \label{KAG2} \ee
Then it follows from (\ref{KAG2}) that the ADM mass of this
solution is $M_K(\infty)=M$, and one can also verify that $M_K(r)$
vanishes at $r=0$. The characteristic behavior of the functions
${\cal E}_e(r)$ and $M_K(r)$ in the vicinity of $r=0$ is shown in
Fig.~2, where we introduced the dimensionless variable $x=Mr/Q^2$
and divided those functions by $M$ for obtaining generic plots not
depending on concrete values of $M$ and $Q$. There, one can
observe the presence of the region with negative values of the
Komar function.

\begin{figure}[htb]
\centerline{\epsfysize=55mm\epsffile{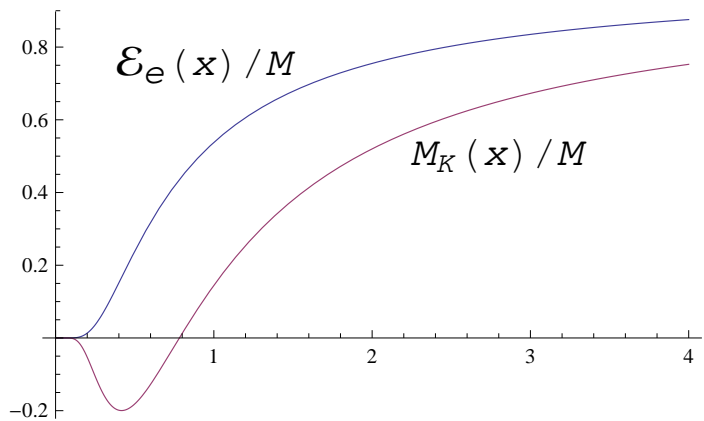}} \caption{Behavior
of the functions ${\cal E}_e(x)/M$ and $M_K(x)/M$, with
$x=Mr/Q^2$, in the case of the second ABG solution for arbitrary
nonzero $M$ and $Q$.}
\end{figure}

\section{The total energy of magnetic field in Bardeen spacetime}

We now turn to analyzing the electromagnetic energy issue in
Bardeen's spacetime given by the metric \cite{Bar}
\be d s^2=-fdt^2+f^{-1}dr^2+r^2d\Omega^2, \quad
f=1-\frac{2Mr^2}{(r^2+Q^2)^{3/2}}, \label{Bar} \ee
in which the parameter $Q$ was originally interpreted as
describing the electric charge, but later reinterpreted by
Ay\'on-Beato and Garc\'ia \cite{AGa3} as representing a nonlinear
magnetic monopole with the electromagnetic tensor
\be F_{\a\beta}=2\delta^\vartheta_{[\a} \delta^\varphi_{\beta]}
Q\sin\vartheta. \label{EP3} \ee

Once again choosing the most convenient way of finding the density
of electromagnetic field solely through the metric (\ref{Bar}), we
readily obtain
\be -T^t_t= \frac{3MQ^2}{4\pi(r^2+Q^2)^{5/2}}, \label{dB} \ee
so that the magnetic energy ${\cal E}_m(r)$ inside a sphere of
radius $r$ will have the form
\be {\cal E}_m(r)=\frac{MR^3}{(r^2+Q^2)^{3/2}}, \label{ER3} \ee
thus leading in the limit $r\to\infty$ to the expectable result
for the total energy of magnetic field:
\be {\cal E}_m(\infty)=M. \label{EB} \ee

As for the Komar mass function associated with the Bardeen
spacetime, it is obtainable from (\ref{Km}), (\ref{wAG1}) and
(\ref{KAG1}) taking into account that
\be \omega_{\vartheta\varphi}
=\frac{2Mr^3(r^2-2Q^2)\sin\vartheta}{(r^2+Q^2)^{5/2}},
\label{wtf3} \ee
hence yielding
\be M_K(r)=\frac{Mr^3(r^2-2Q^2)}{(r^2+Q^2)^{5/2}}. \label{KB} \ee
As a result, the ADM mass $M_K(\infty)$ of this spacetime is equal
to $M$, similar to the two ABG solutions previously considered.

It follows from (\ref{KB}) that the Komar function of Bardeen's
model takes negative values on the interval $0<r<\sqrt{2}|Q|$ (of
course, we assume that $M>0$), and it has one minimum at
$r_m=\sqrt{\frac23}|Q|$, so that for $r>r_m$, $M_K(r)$ is an
increasing function. Note that although the functions ${\cal
E}_m(r)$ and $M_K(r)$ in (\ref{ER3}) and (\ref{KB}) differ from
the respective expressions in the ABG solutions, still their
behavior in Bardeen's case depicted in Fig.~3 is very similar to
that shown earlier in Figs.~1 and 2.

\begin{figure}[htb]
\centerline{\epsfysize=55mm\epsffile{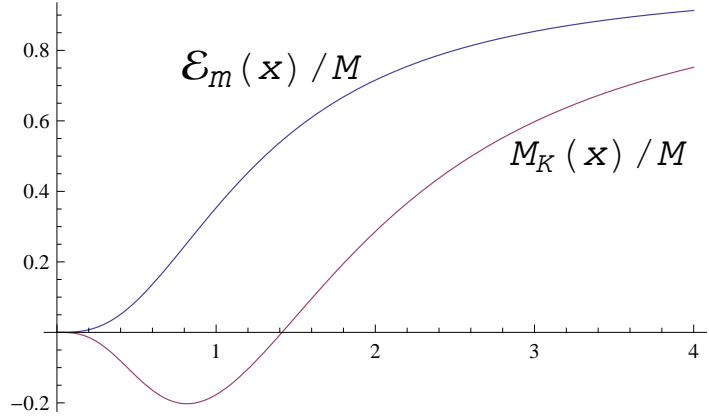}} \caption{Behavior
of the functions ${\cal E}_m(x)/M$ and $M_K(x)/M$, with $x=r/|Q|$,
in the case of the Bardeen spacetime for arbitrary nonzero $M$ and
$Q$.}
\end{figure}

It is worth mentioning that in the case of the Schwarzschild
solution the component $\omega_{\vartheta\varphi}$ of the 2-form
$\wb$ is equal to $2M\sin\vartheta$, being independent of $r$, and
consequently $M_K(r)=M$ for any $r>0$, which means that the whole
mass of the Schwarzschild black hole is contained in the
singularity at $r=0$. In this respect, it appears that the mass in
the Bardeen and ABG models is not localized in some restricted
region but rather is distributed over the entire space.

Let us also note for completeness that in the Reissner-Nordstr\"om
solution, which is singular at $r=0$, the expression for the
density of electric field does not involve the mass parameter $M$,
being equal to $Q^2/8\pi r^4$. This implies that the corresponding
expression of the electric energy is independent of $M$ too; and
although (as is well known) the respective integral over the whole
space is divergent, still the integration over $r$ makes sense on
the interval $[r,+\infty)$, $r>0$, giving $Q^2/2r$. The analogous
energy of magnetic field in Bardeen's model on the latter interval
is equal to $M-Mr^3(r^2+Q^2)^{-3/2}$, and it vanishes when either
of the parameters $M$ or $Q$ is equal to zero.

\section{Concluding remarks}

It is really surprising that all three different models of
non-singular black-hole spacetimes considered in the present paper
share the same fundamental characteristic with regard to the issue
of the total electromagnetic energy whose value, on the one hand,
turns out to be independent of the charge parameter $Q$ and, on
the other hand, is equal exactly to the ADM mass $M$. At the same
time, this result strongly suggests that, from the global point of
view, the entire ``mass'' in the Bardeen and ABG models comes from
the electromagnetic field and the particular values of $Q$ do not
affect it. Indeed, after converting the Schwarzschild singularity
(that contained the whole mass) into a regular mass distribution
by means of nonlinear electrodynamics, one is obliged to explain
the origin of that novel mass distribution through the
corresponding energy-momentum tensor. So, when the latter tensor
is that of the electromagnetic field only, with no any other
sources of gravity, then one inevitably arrives at the conclusion
that the mass in such regular spacetimes must have the
electromagnetic origin. In this respect, it would be worth
recalling the original paper of Born and Infeld \cite{BIn} in
which the modified Maxwell's equations had been used for deducing
the electromagnetic origin of inertia, and we have an impression
that in the papers \cite{AGa1,AGa2,AGa3} this old idea
contradicting the modern conception about the nature of mass was
just reproduced at a new level.

\section*{Acknowledgments}

This work was partially supported by Project~FIS2012-30926 from
MCyT of Spain, and by CONACYT of Mexico. One of us (ER) thanks
M\'aximo L\'opez, Head of the Physics Department of CINVESTAV,
where part of this work was done, for the hospitality extended to
him during his visit.





\bibliographystyle{model1a-num-names}
\bibliography{<your-bib-database>}

\begin{thebibliography}{100}


\bibitem{Bar} J. Bardeen,
in Abstracts of GR5 International Conference (Tbilisi, 1968).

\bibitem{Rei} H.~Reissner, \J{\APL}{355}{1916}{106}.

\bibitem{Nor} G.~Nordstr\"om, Proc. K. Ned. Akad. Wet. 20 (1918) 1238.

\bibitem{Bor} A. Borde, \J{\PRD}{50}{1994}{3692}.

\bibitem{AGa1} E. Ay\'on-Beato, A. Garc\'ia, \J{\PRL}{80}{1998}{5056}.

\bibitem{AGa2} E. Ay\'on-Beato, A. Garc\'ia, \J{\PLB}{464}{1999}{25}.

\bibitem{AGa3} E. Ay\'on-Beato, A. Garc\'ia, \J{\PLB}{493}{2000}{149}.

\bibitem{Agu} J.M. Aguirregabir\'ia, Computer code RICCI, 2002.

\bibitem{ADM} R. Arnowitt, S. Deser, C.W. Misner,
\J{\PR}{122}{1961}{997}.

\bibitem{Kom} A.~Komar, \J{\PR}{113}{1959}{934}.

\bibitem{BIn} M. Born, L. Infeld, \J{Nature}{132}{1933}{970}.

\end{thebibliography}



\end{document}